\pdfoutput=1
\documentclass[11pt,a4paper,final]{article}
\usepackage[left=1in,right=1in,top=1in,bottom=1in]{geometry}

\usepackage{microtype}
\usepackage{graphicx}
\usepackage{booktabs} 
\usepackage{hyperref}
\usepackage[numbers]{natbib}
\usepackage{authblk}
\usepackage{tikz}

\usepackage{pifont}
\usepackage{booktabs}
\usepackage{longtable}

\usepackage{amsmath}
\usepackage{amssymb}
\usepackage{mathtools}
\usepackage{amsthm}

\usepackage[capitalize,noabbrev]{cleveref}
\usepackage{tabularx}
\usepackage{enumitem}
\usepackage{multirow}
\usepackage{comment}
\usepackage{color-edits}
\addauthor{vc}{blue}
\usepackage{bbm}

\usepackage{listings}
\usepackage{xcolor}
\usepackage{xspace}
\usepackage{pifont}
\usepackage{mdframed}
\usepackage[most]{tcolorbox}
\DeclareRobustCommand{\halfcheckmark}{%
  \protect\tikz[]\protect\draw[scale=0.3,fill=black]
    (0,.35) -- (.25,0) -- (1,.7) -- (.25,.15) -- cycle
    (0.75,0.2) -- (0.77,0.2) -- (0.6,0.7) -- cycle;%
}

\newtcolorbox{examplebox}[1][]{
    enhanced,
    colback=white,
    colframe=black,
    boxrule=0.5pt,
    fontupper=\small\ttfamily,
    arc=0mm,
    outer arc=0mm,
    #1
}

\newtcbtheorem{Summary}{\bfseries Takeaway}{enhanced,drop shadow={black!50!white},
  coltitle=black,
  top=0.15in,
  attach boxed title to top left=
  {xshift=1.5em,yshift=-\tcboxedtitleheight/2},
  boxed title style={size=small,colback=white}
}{summary}

\newtcolorbox[auto counter]{summary}[1][]{title={\bfseries~#1 Findings},enhanced,drop shadow={black!50!white},
  coltitle=black,
  top=0.1in,
  attach boxed title to top left=
  {xshift=1.5em,yshift=-\tcboxedtitleheight/2},
  boxed title style={size=small,colback=white},}

\usepackage{xcolor}
\usepackage{xspace}
\usepackage[most]{tcolorbox}
\usepackage{fontawesome}
\usepackage{pgfplots}
\usepackage{pgfplotstable}
\usepackage{multirow}
\usepackage{multicol}

\theoremstyle{plain}

\theoremstyle{definition}

\theoremstyle{remark}

\usepackage{caption}
\usepackage{subcaption}

\definecolor{green3}{RGB}{66,179,130}
\definecolor{green2}{RGB}{121,205,169}
\definecolor{green1}{RGB}{196,233,217}

\definecolor{red1}{RGB}{251,219,220}
\definecolor{red2}{RGB}{244,164,166}
\definecolor{red3}{RGB}{236,91,96}

\definecolor{blue1}{RGB}{101,173,246}
\definecolor{blue2}{RGB}{12,112,212}

\definecolor{orange1}{RGB}{250,181,97}
\definecolor{orange2}{RGB}{245,138,7}

\definecolor{purple1}{RGB}{195,176,232}
\definecolor{purple2}{RGB}{146,112,212}

\definecolor{pink1}{RGB}{236,152,223}
\definecolor{pink2}{RGB}{227,100,208}

\definecolor{gray1}{RGB}{220,220,220}

\def\importancebarchart#1#2#3#4#5#6#7#8{
\resizebox{0.08\linewidth}{7.5pt} {
\begin{tikzpicture}[]
\node[] { \huge \emph{#7}};
\end{tikzpicture}
}
\resizebox {0.81\linewidth} {6.5pt} {%
\begin{tikzpicture}[]
\begin{axis}[
      axis background/.style={fill=gray!30, draw=gray!30},
      axis line style={draw=none},
      tick style={draw=none},
      ytick=\empty,
      xtick=\empty,
      ymin=0, ymax=0.70,
      xmin=0, xmax=6]
\addplot [
      ybar interval=.5,
      fill=blue2,
      draw=none,
]
	coordinates {(6*#1,1) (0,0.30)}; %
\addplot [
      ybar interval=.5,
      fill=blue1,
      draw=none,
]
	coordinates {(6*(#1+#2),1) (6*#1,1)}; %
\addplot [
      ybar interval=.5,
      fill=gray1,
      draw=none,
]
	coordinates {(6*(#3+#2+#1),1) (6*(#2+#1),1)}; %

\addplot [
      ybar interval=.5,
      fill=orange1,
      draw=none,
]
	coordinates {(6*(#4+#3+#2+#1),1) (6*(#3+#2+#1),1)}; %
\addplot [
      ybar interval=.5,
      fill=orange2,
      draw=none,
]
	coordinates {(6*(#5+#4+#3+#2+#1),1) (6*(#4+#3+#2+#1),1)}; %
\addplot [
      ybar interval=.5,
      fill=orange2,
      draw=none,
]
	coordinates {(6*(#6+#5+#4+#3+#2+#1),1) (6*(#5+#4+#3+#2+#1),1)}; %
\end{axis}%
\end{tikzpicture}%
}
\resizebox{0.08\linewidth}{7.5pt} {
\begin{tikzpicture}[]
\node[] { \huge \emph{#8}};
\end{tikzpicture}
}
}

\def\mylegend#1#2{
\resizebox {0.02\linewidth} {6.5pt} {%
\begin{tikzpicture}[]
\begin{axis}[
      axis background/.style={fill=white!30, draw=white!30},
      axis line style={draw=none},
      tick style={draw=none},
      ytick=\empty,
      xtick=\empty,
      ymin=0, ymax=0.70,
      xmin=0, xmax=6]
\addplot [
      ybar interval=.5,
      fill=#2,
      draw=none,
]
	coordinates {(4.5,1) (0,0.30)}; %
\end{axis}%
\end{tikzpicture}%
}%
#1
}

\definecolor{boxcolor}{RGB}{238, 223, 204} %

\allowdisplaybreaks

\title{\textbf{Code with Me or for Me? How Increasing AI Automation Transforms Developer Workflows}}

\renewcommand{\thefootnote}{\fnsymbol{footnote}}

\begin{document}

\author[1,2]{Valerie Chen}
\author[1]{Ameet Talwalkar}
\author[2]{Robert Brennan}
\author[1,2]{Graham Neubig}
\affil[1]{Carnegie Mellon University}
\affil[2]{All Hands AI}

\date{Correspondence to \texttt{valeriechen@cmu.edu}}

\maketitle

\renewcommand*{\thefootnote}{\arabic{footnote}}

\setcounter{footnote}{0}

\begin{abstract}
Developers now have access to a growing array of \emph{increasingly autonomous} AI tools for software development.
While many studies examine copilots that provide chat assistance or code completions, evaluations of coding agents—which can automatically write files and run code—still rely on static benchmarks.
We present the first controlled study of developer interactions with coding agents, characterizing how more autonomous AI tools affect productivity and experience.
We evaluate two leading copilot and agentic coding assistants, recruiting participants who regularly use the former.
Our results show agents can assist developers in ways that surpass copilots (e.g., completing tasks humans may not have accomplished) and reduce the effort required to finish tasks.
Yet challenges remain for broader adoption, including ensuring users adequately understand agent behaviors.
Our findings reveal how workflows shift with coding agents and how interactions differ from copilots, motivating recommendations for researchers and highlighting challenges in adopting agentic systems.
\end{abstract}

\begin{figure}
\includegraphics[width=\textwidth]{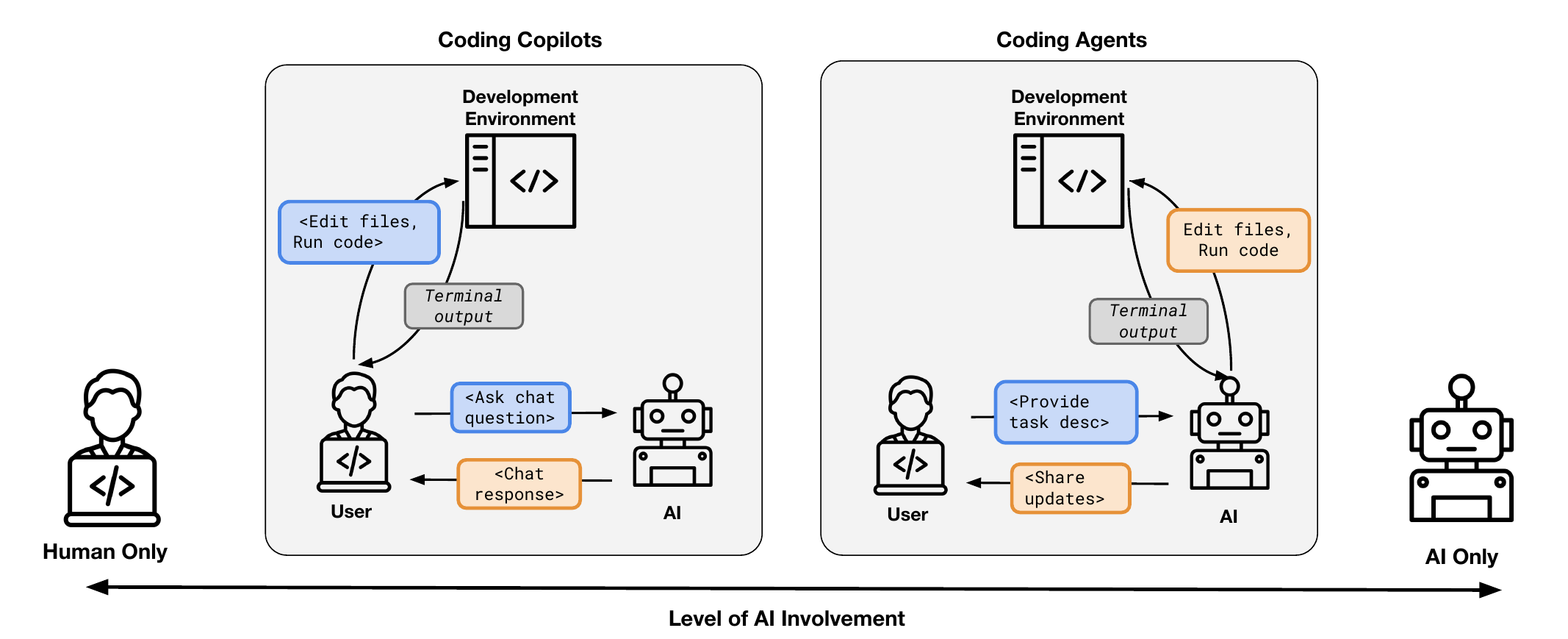}
\caption{\textbf{The spectrum of developer workflows with AI tools.}
While prior studies have evaluated how human-only workflows compare to human-copilot workflows, understanding how AI can code \emph{with} developers, little work exists comparing copilots to the emerging agent workflows, where AI is more often used to code \emph{for} developers.
We study how these differences affect productivity, user experience, and interaction patterns with AI tools.}
\label{fig:tools}
\end{figure}

\section{Introduction}

AI-powered tools have become integral to modern software development workflows. Tools such as GitHub Copilot~\citep{copilot}, Cursor~\citep{cursor}, Windsurf~\citep{windsurf}, and Continue~\citep{continue_2025} primarily function as \textit{copilots}---offering autocomplete suggestions and chat-based code assistance. 
These systems assist developers incrementally, typically requiring continuous developer input to drive progress by making further changes themselves within the development environment.
In contrast, coding \textit{agents} such as Devin~\citep{devin}, OpenHands~\citep{wang2024openhands}, and Claude Code~\citep{claudecode} are designed for greater autonomy: they can autonomously plan multi-step actions based on user inputs, navigate across files, execute code, and iteratively revise outputs.
One might draw a parallel where coding with copilots is akin to a pair programming~\citep{begel2008pair} whereas collaborating with coding agent might be more akin to working with a fellow engineer.
Though some have claimed that coding agents will be eventual replacements for human developers~\citep{replace}, until they demonstrate a high level of reliability across diverse contexts, these systems will necessarily require close human supervision.
As developers gain access to increasingly autonomous AI tools (Figure~\ref{fig:tools}), it is critical to understand how these systems are transforming software workflows, developer productivity, and overall user experience.

Despite growing commercial interest in agents for software development, understanding how well coding agents perform in real-world settings with software engineers remains incomplete for multiple reasons.
First, the AI community has largely relied on static benchmarks to evaluate new agents~\citep{swe-bench,yang2024swe,zan2025multi}, which have well-specified problems and no human-in-the-loop to provide interactive feedback.
While initial work aims to incorporate interactivity with synthetic users on benchmark tasks~\citep{shao2024collaborative,vijayvargiya2025interactive}, synthetic users struggle to capture all the nuances of real users~\citep{prpa2024challenges,hwang2025human}.
In contrast, there is a much richer body of work on understanding how developers use various copilot features~\citep{vaithilingam2022expectation,peng2023impact,barke2022grounded,prather2023its, mozannar2022reading} and even designing new affordances to improve interactions with them~\citep{yan2024ivie,vasconcelos2023generation,chen2024need}.
Prior works examining how developers use existing copilots have even demonstrated that benchmark performance alone may not necessarily correlate with downstream user utility~\citep{mozannar2024realhumaneval}.
Second, no prior work has considered a head-to-head comparison of copilots and agents, despite developers often having access to both types of tools in their day-to-day.
Without controlled comparisons with existing tools, it remains unclear how much current agentic systems improve software development.

In this work, we conduct the first controlled user study to understand how developers interact with coding agents and compare them to existing copilots.
In particular, we recruit a population of users who are experienced copilot users but novice agent users, which is representative of the majority of current software developers according to a 2025 Stack Overflow developer survey~\citep{stack_overflow_survey}.
We adopt a within-participant design where participants solve realistic coding problems with \textit{both} Github Copilot~\citep{copilot}---one of the most widely used AI copilots for code---and OpenHands~\citep{wang2024openhands}---a state-of-the-art, open-source coding agent.
Each respective choice encompasses many common features of other popular copilot or agent alternatives.
Additionally, we ask participants to solve tasks that are motivated by realistic use cases~\citep{mialon2023gaia,swe-bench}, which include creating programs from scratch, implementing features, and fixing bugs in real-world repositories. 
Since participants work on tasks of the same type with both the copilot and coding agent, we can directly compare their impacts on user productivity and user experience.
Furthermore, by indirectly monitoring participant interaction with both types of AI tools as they are solving the problem, we can also derive a better understanding of interaction patterns across copilots and agents.

We find that, from the productivity angle, agents tend to enable users to complete more tasks than with copilots, observing an average $35\%$ increase in task correctness, and with about $50\%$  the user effort, as measured by the amount of time spent by the user.
On the user experience front, we also observe that this result translates into significantly less cognitive load and also a significant improvement in a user's sense of ability to accomplish new tasks.
When more carefully inspecting interaction trajectories with each type of tool, we find that the reduction in user effort is largely due to the agent's ability to autonomously debug and set up complex environments and assist users in evaluating whether code implementations follow task instructions.
However, despite productivity gains, participants noted multiple aspects of the user experience could be improved, including user satisfaction and understanding of agent outputs.  
We discuss three desiderata for building more collaborative coding agents.

Together, our approach and findings offer multiple contributions: 
(1) Our study tackles the timely and important issue of understanding the effect of increasing automation of AI coding assistants. We do this through a study design that directly contrasts how developers use copilots compared to coding agents. 
(2) Through our study, we show that coding agents can have substantial productivity implications on modern software development. Our findings, however, caution against the immediate adoption of existing tools given limitations in user experience. 
(3) We provide an in-depth discussion of the implications of our findings, limitations of our work, and future research directions. 
In particular, we identify three goals for building future coding agents that include improved transparency, balanced proactivity, and effective involvement of human effort.
While the focus of this work is on agents in the context of software engineering tasks, we believe our findings and recommendations can be more broadly applicable to designing better emerging human-agent interactions.

\begin{figure}[t]
\centering
\includegraphics[width=0.8\textwidth]{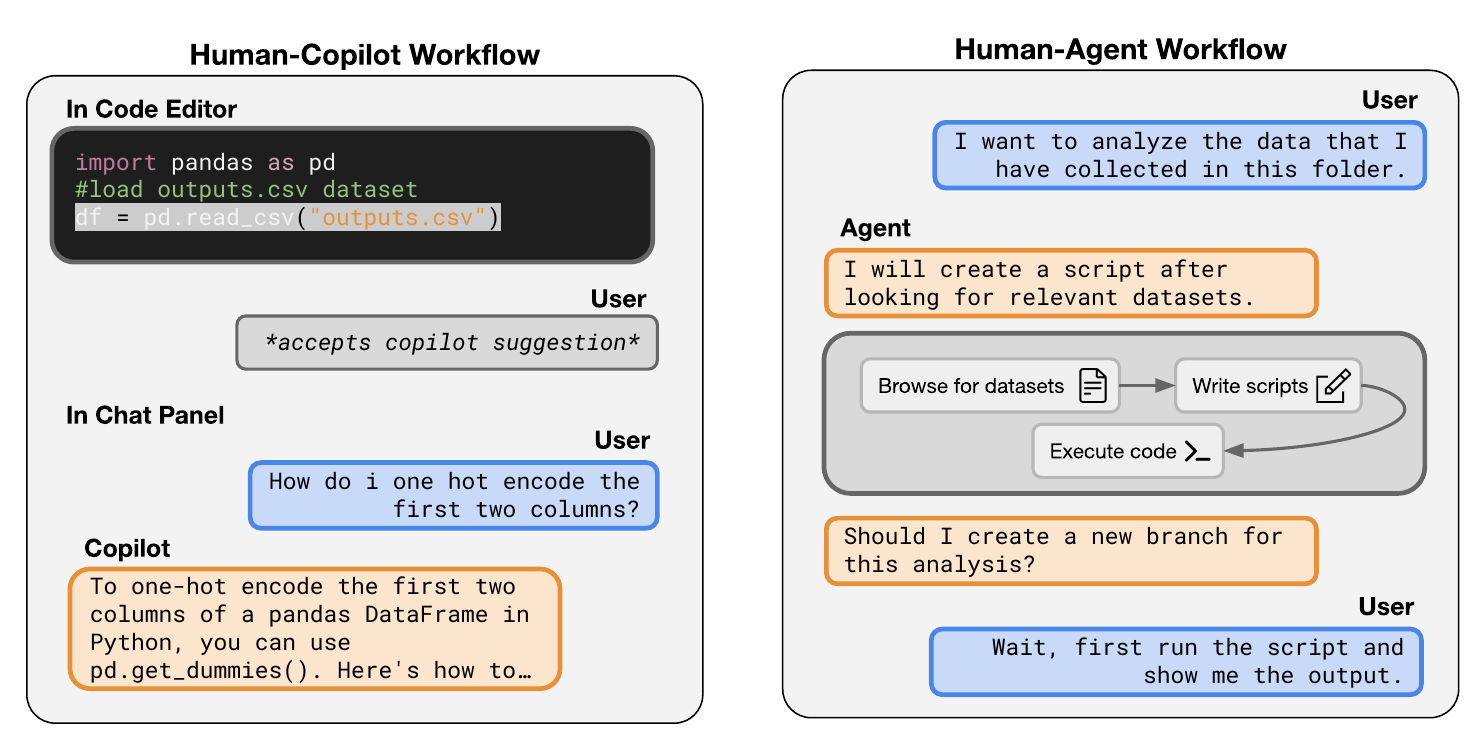}
\caption{\textbf{Comparing affordances in copilot (left) and agent (right) workflows.}
While prior work has studied the benefits and challenges of coding with copilots (overviewed in Section~\ref{subsec:workflow}), we ask the same questions about agentic workflows.}
\label{fig:interface_comparison}
\end{figure}

\section{Related Work}

\subsection{The Ecosystem of AI Coding Assistants}

A growing number of tools powered by large language models (LLMs)---i.e., AI coding assistants---are available to developers to generate or edit code and answer queries. 
Developers are increasingly writing code with AI assistants like Github Copilot~\cite{copilot} and Cursor~\cite{cursor} and are using chat assistants like ChatGPT~\cite{chatgpt} or Claude~\cite{claude} in place of online Q\&A communities like Stack Overflow~\cite{xiao2023devgpt}.
We refer to these types of coding assistants as \emph{copilots} because these are tools that assist the developer but are not in command.
Copilots typically involve one of two types of support (Figure~\ref{fig:interface_comparison}, left): autocomplete suggestions are used to quickly write more code based on the programmer's current code context, while chat dialogue can help answer questions that range from debugging errors to explaining documentation.

More recently, we have seen researchers and companies introduce coding \emph{agents} (e.g., Devin~\citep{devin}, OpenHands~\citep{wang2024openhands}, Claude Code~\citep{claudecode}).
This is largely due to recent advances in LLM capabilities, which have sparked a surge in the development of AI agents~\citep{wu2023autogen, weng2023agent, Sumers2023CognitiveAF, durante2024interactive}.
We refer to AI agents as LLM-powered systems that are equipped with the ability to use tools, run commands, observe feedback from the environment, and plan for future actions.
In the context of software engineering tasks, unlike copilots, coding agents can autonomously execute the same set of actions that a human developer can, provided a natural language task description from the developer (Figure~\ref{fig:interface_comparison}, right).
Note that coding agents typically do not have autocomplete functionality, as that is considered too limited in scope.
Given the increased interest in agents, there is a need to understand how coding agents compare to existing tools like copilots.
In this work, we conduct a study that compares a widely used copilot (e.g., GitHub Copilot~\citep{copilot}) to a state-of-the-art coding agent (e.g., OpenHands~\cite{wang2024openhands}).

\subsection{Impact of AI Coding Assistants In Practice}

To assess whether a new AI tool can be integrated into a developer's workflow effectively, multiple factors should be assessed, including the performance of the tool on its own (or the model powering this tool), the performance on human and AI compared to other alternatives, the human's experience, and so on.  
We discuss how these various factors have been studied by HCI and AI communities in the context of both coding copilots and agents:

Evaluations of copilot capabilities began by measuring an LLM's performance on coding benchmarks that test the model's code generation capabilities (e.g., HumanEval~\citep{chen2021evaluating}). 
As LLM performance on static benchmarks improved, a growing and fairly extensive set of user studies has been conducted to understand how developers code with various AI copilots.
These studies have focused on two forms of  support present in AI copilots: autocomplete 
suggestions~\citep{vaithilingam2022expectation,peng2023impact,barke2022grounded,prather2023its, mozannar2022reading, vasconcelos2023generation,cui2024productivity,mozannar2024realhumaneval} and chat dialogue~\citep{ross2023programmer, chopra2023conversational, kazemitabaar2023studying, gu2023analysts,nam2024using,mozannar2024realhumaneval}.
These studies show how copilots have generally had a positive impact on software development, e.g., leading to an increase in perceived productivity~\citep{ziegler2022productivity,10.1145/3706599.3706670,becker2025measuring} and rate of task completion in controlled studies~\citep{vaithilingam2022expectation, peng2023impact} compared to developers writing code on their own.

Given that coding agents are a much more nascent technology, attempts to measure their usefulness on software engineering tasks have largely been limited to benchmarks without humans-in-the-loop~\citep{swe-bench,yang2024swe,zan2025multi, shao2024collaborative,vijayvargiya2025interactive}, which cannot provide insight into the impacts on user experience and productivity.
Recent efforts track the usage of the various aforementioned coding agents on GitHub~\citep{aitw2025dashboard,li2025rise} indicate a growing trend in adoption.
Initial analysis of this data suggests that coding agents may still lag behind humans on PR acceptance rates by a large margin, especially on complex tasks.
Our work complements these analyses by quantifying, on a set of coding tasks, the difference in productivity and user experience, while identifying opportunities for the community to design better agents.

\subsection{Developer Workflows with AI Tools}\label{subsec:workflow}

The growing use of AI coding assistants has motivated studies towards understanding developer workflows with AI tools---i.e., where do people tend to spend their time when working with AI coding assistants, and what are the bottlenecks created in these workflows.
We focus our discussion on prior studies of copilot systems, as that is where the primary set of related work lies.
The focus of multiple works has been in the autocomplete setting~\citep{barke2022grounded,mozannar2022reading}.
\citet{mozannar2022reading} developed CUPS, a taxonomy of common programmer activities when interacting with code completions, and observed that ``verifying suggestions'' takes up a significant portion of a developer's AI workflow.
Another work by~\citet{barke2022grounded} introduced the terms ``acceleration'' and ``exploration'' to denote the difference between cases where code completion models are used to help developers accomplish something more quickly versus explore potential solutions, indicating two different roles of AI tools.

As code completion is no longer the only point of interaction with AI coding assistants, recent work has shifted towards understanding natural language-driven interaction in chat interfaces.
Real conversations between developers and AI show the diversity~\cite{xiao2023devgpt}, from explaining existing code, brainstorming new ideas or functionality, to identifying and fixing bugs.
Within this interaction format, prior work has found that one bottleneck is the amount of time spent constructing prompts~\citep{chopra2023conversational,nam2024using}.
In the context of agents, some may associate coding with agents with the growing paradigm of ``vibe-coding,'' where developers tend to write natural language task descriptions rather than the code itself~\citep{sarkar2025vibe}.
Relatedly,~\citet{10.1145/3706598.3714154} studied a complementary direction of supporting exploration of program design space with more agentic workflows.
Our work focuses on the implementation process that requires iterating on code and debugging once a task description is provided. 
We will later revisit some of these observations across autocomplete and chat to discuss how they compare to interactions with agents.

% \vccomment{do we want to mention scientific, data analysis workflows?}

\section{Study Design}

As developers begin to adopt coding agents, we aim to understand when and by how much better agents are than existing copilots and investigate how people interact with coding agents. Our research questions are as follows: 

\begin{itemize}[leftmargin=.5in]
    \item[\textbf{RQ1:}] \textit{How do agents and copilots compare in terms of effect on user productivity?} While agents are often marketed as AI software engineers, we evaluate to what extent they can help a user complete more tasks than when a user has access to a widely used copilot tool. 
    \item[\textbf{RQ2:}] \textit{How do agents and copilots compare in terms of effect on user experience?} Extensive surveys have been conducted to understand user experience with copilots. We ask many of the same questions that have been explored in prior surveys but directly comparing agents and copilots.
    \item[\textbf{RQ3:}] \textit{How do developers interact with different coding assistants?} Prior work has explored modeling user behaviors with copilots~\citep{mozannar2022reading,barke2022grounded,peng2023impact}. We compare interaction patterns between both tools. 

\end{itemize}

\subsection{Participants}

We recruited a total of 20 students via university mailing lists.
The inclusion criteria for the study are that they must have access to and use GitHub Copilot regularly and have experience programming in Python.
Since we select tasks that are largely written in Python, we require participants to have a baseline level of Python knowledge.
Among our participants, 60\% had 3-5 years of professional programming experience, 25\% had 0-2 years, and 15\% had 6-10 years.
Additionally, 95\% of participants rated themselves as having intermediate to advanced proficiency in Python.
In terms of AI tool usage, all participants used GitHub Copilot at least once a week in addition to other common chat assistants like ChatGPT, Claude, and Gemini.
Unsurprisingly, no participants reported any prior experience with using coding agents, as current usage of agents largely remains in communities of early adopters~\citep{stack_overflow_survey}. 
Accordingly, our results are targeted towards the broader set of developers who are still largely using copilots, and we aim to understand how they will adopt this new technology.

\begin{table}[t]
\centering
\resizebox{\textwidth}{!}{%
\begin{tabular}{l|cc|cccc}
\toprule
\multirow{2}{*}{\textbf{Tool}} & \multicolumn{2}{c|}{\textbf{Basic Features}} & \multicolumn{4}{c}{\textbf{Advanced Capabilities}} \\
\cmidrule(lr){2-3} \cmidrule(lr){4-7}
& Auto- & Chat & Multi-file & Terminal & File & Act  \\
& complete & & Editing & Execution & Creation & Autonomously \\
\midrule
\multicolumn{7}{l}{\textbf{Copilots}} \\
\midrule
GitHub Copilot*~\citep{copilot} & $\checkmark$ & $\checkmark$ & $\halfcheckmark$ & $\halfcheckmark$ & $\halfcheckmark$ & $\times$ \\
Cursor~\cite{cursor} & $\checkmark$ & $\checkmark$ & $\halfcheckmark$ & $\halfcheckmark$ & $\halfcheckmark$ & $\times$ \\
Windsurf~\cite{windsurf} & $\checkmark$ & $\checkmark$ & $\halfcheckmark$ & $\times$ & $\times$ & $\times$ \\
Continue~\cite{continue_2025} & $\checkmark$ & $\checkmark$ & $\times$ & $\times$ & $\times$ & $\times$ \\
Amazon CodeWhisperer~\citep{codwhisperer} & $\checkmark$ & $\checkmark$ & $\times$ & $\times$ & $\times$ & $\times$ \\
\midrule
\multicolumn{7}{l}{\textbf{Agents}} \\
\midrule
OpenHands*~\cite{wang2024openhands} & $\times$ & $\checkmark$ & $\checkmark$ & $\checkmark$ & $\checkmark$ & $\checkmark$ \\
Devin~\cite{devin} & $\times$ & $\checkmark$ & $\checkmark$ & $\checkmark$ & $\checkmark$ & $\checkmark$ \\
Claude Code~\cite{claudecode} & $\times$ & $\checkmark$ & $\checkmark$ & $\checkmark$ & $\checkmark$ & $\checkmark$ \\
Cline~\cite{cline} & $\times$ & $\checkmark$ & $\checkmark$ & $\checkmark$ & $\checkmark$ & $\checkmark$ \\
Replit~\cite{replit} & $\times$ & $\checkmark$ & $\checkmark$ & $\times$ & $\checkmark$ & $\checkmark$ \\
\bottomrule
\end{tabular}
}
\caption{\textbf{Comparison of popular AI coding copilots and agents}, where $\checkmark$ denotes the action is done by the AI, $\halfcheckmark$ denotes the action is approved by a human, and $\times$ denotes the feature does not exist. We split features based on basic ones that involve the current file and more advanced capabilities. We use * to denote the choice of copilot and agent that we select for this study.}
\label{tab:ai_coding_tools}
\end{table}

\subsection{Choice of Copilot, Agent, and LLMs}\label{subsec:choices}

We discuss our choice of copilot and agent for the study, which encompasses many relevant functionalities from popular alternatives (Table~\ref{tab:ai_coding_tools}).
Additional details about how users would interact with each are provided in the Appendix~\ref{appdx:interfaces}

\paragraph{Choice of copilot.} 
GitHub Copilot~\citep{copilot} is one of the most widely used AI coding assistants, making it a natural choice for the copilot baseline. 
Its popularity also makes it relatively easier to recruit participants who already have access to this tool.
In fact, in our recruitment survey, it was by far the most mentioned AI tool that participants regularly used.
Furthermore, it has been the sole choice of AI coding assistant in the majority of prior studies aiming to understand developer usage of AI tools~\citep{vaithilingam2022expectation,peng2023impact,barke2022grounded,prather2023its, mozannar2022reading, vasconcelos2023generation}.
Despite choosing one instantiation of coding copilots, GitHub Copilot is also representative of other tools like Cursor, Windsurf, and Continue: it can generate code completions based on the user's cursor position, answer user questions in the chat panel, incorporate suggested code edits into the user's file, or suggest code edits in a diff format for users to review. 
GitHub Copilot is a tool comprised of many LLMs---a faster, light-weight code generation model and a flexible choice of models that powers the chat assistant.
In the copilot portion of the study, we allow participants to use any LLM they want in the chat panel to most closely mimic real-world usage, where some users may be more familiar with some LLMs.
At the time of the study, the LLMs that are available to users include \texttt{GPT-4o}, \texttt{Claude Sonnet 3.7}, \texttt{Gemini 2.5 Pro}, and \texttt{o3-mini}.

\paragraph{Choice of agent.} 
Given that agents are much newer on the market, there is not yet one option that has been widely adopted by developers.
For our study, we select OpenHands, which is a leading open-source coding agent on benchmarks like SWE-Bench~\citep{swe-bench}.
OpenHands supports multiple tools, including a bash shell, Jupyter IPython server, text-based browsing tool that uses a Chromium browser, and a file-processing tool for creating, viewing, and editing plain-text files.
Users primarily interact with OpenHands through a chat interface, but can use tabs on the side of the interface to see an overview of changes, modify files themselves through an embedded VSCode interface, view web servers or browser through a non-interactive.
In contrast to copilots, agents are typically optimized for specific LLM backbones.
As such, we only allow users to interact with a version of OpenHands powered by \texttt{Claude Sonnet 3.7}.
Since the models that participants have access to in the copilot phase are regarded as being similarly proficient as the fixed model, we consider it to be a reasonable setup. 
We later evaluate whether this limitation impacts user productivity.

\begin{figure*}[t]
\centering
\includegraphics[width=0.85\textwidth]{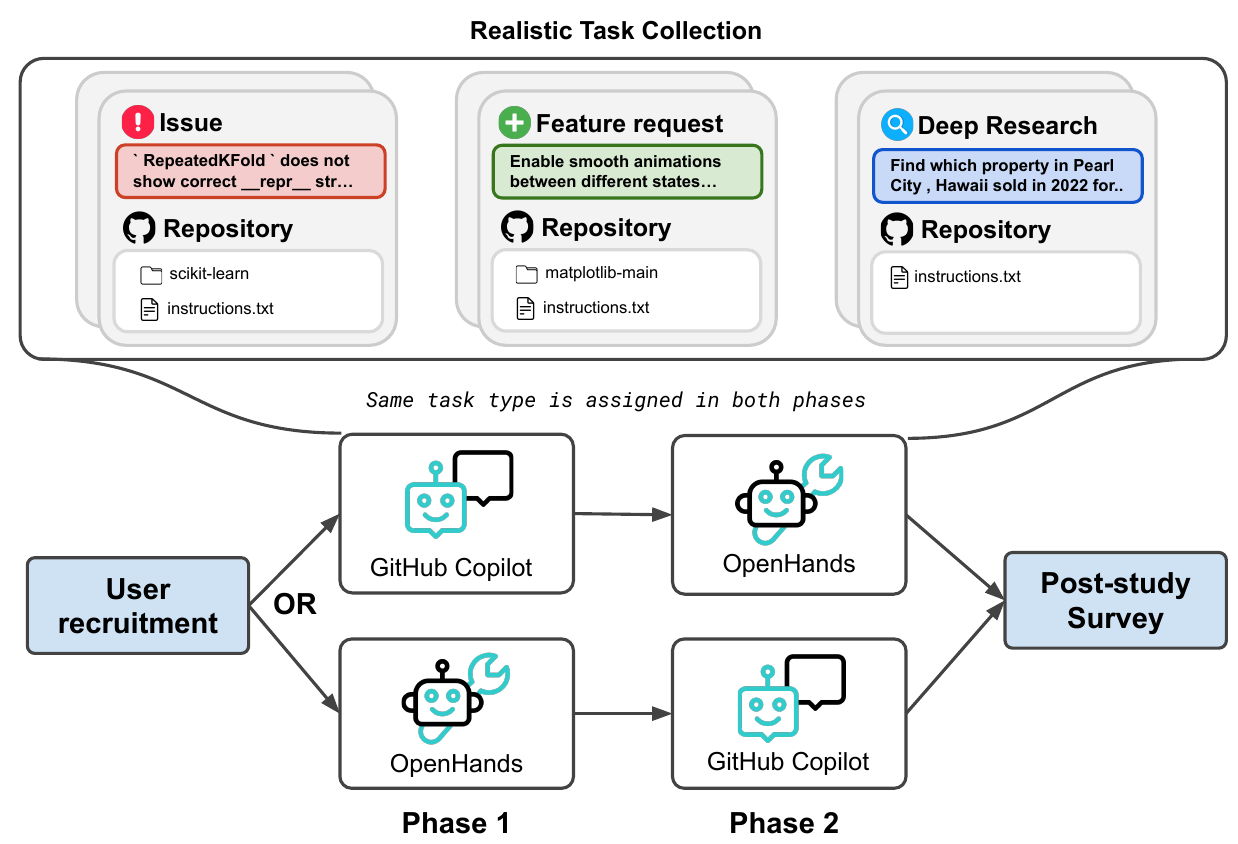}
\caption{\textbf{Overview of our study procedure.} We designed a controlled user study to directly contrast participant interactions with copilots and agents. Participants were asked to solve a set of realistic and challenging tasks, including fixing issues in repositories, adding features, and conducting deep research and analysis.}
\label{fig:study_overview}
\end{figure*}

\subsection{Task Types}

Prior studies evaluating interactions with AI coding assistants used relatively simple tasks that involved only a single file or unrealistic tasks like programming interview-style programs~\citep{vaithilingam2022expectation,mozannar2022reading,chen2024need}. 
We address these limitations by focusing on more challenging tasks, of 3 types: (1) deep research and analysis---which involves researching and scraping data from the internet and writing scripts to process the data, (2) adding features to an existing repository, and (3) fixing bugs in an existing repository.
We source tasks from existing datasets that are used to evaluate agents and select example questions that are not solved by existing agents, and the task description itself is accessible to a broad set of potential participants. 
For (1), we select questions from  GAIA~\citep{mialon2023gaia}, which evaluates AI agents as general-purpose assistants using tasks that require browsing the open web, performing web search, coding, and processing multimodal content. 
The coding tasks in GAIA generally require the agent to write and execute stand-alone programs from scratch. 
For (2) and (3), we select tasks from SWE-Bench~\citep{swe-bench}, which require users to fix issues by editing \emph{existing} code files in a larger, real-world repository (e.g., \texttt{matplotlib} and \texttt{scikit-learn}). 
While SWE-Bench does not directly contain questions that test (2), we manually create variants of the existing questions that fit the feature addition category. 
In total, we had 2 questions per task type.
We provide a list of all tasks in Appendix~\ref{appdx:task_details}.

\subsection{Procedure}

We conducted this study in April-May 2025 following the study procedure outlined in Figure~\ref{fig:study_overview}.
We adopt a within-participant setup, where each participant interacts with both OpenHands and GitHub Copilot.
We randomize the order in which participants interact with either.
The total amount of time the participant spent coding for the study is 90 minutes, with participants spending 40 minutes in each condition and additional time for onboarding and filling out the post-task survey.
Participants conduct the study asynchronously in their own time.
Each participant is sent a document outlining the study procedure, which also contains links to GitHub repositories where they can download relevant starter code for each task.
Upon completion, participants were asked to push their updated code to a new repository to be shared with the first author and were subsequently compensated with a \$35 Amazon gift card.

We use a between-subjects set-up to randomize task types.
This means that in both phases,  participants would work on the \textit{same} type of task using both Copilot and OpenHands.
This not only allows us to control for problem difficulty but also allows participants to more directly contrast their experience working on the same type of task with two different AI tools.
Finally, we ask participants to record themselves when writing code with GitHub Copilot. 
Since participants already have access to GitHub Copilot, there is little setup required to get started on this portion of the study.
When working with OpenHands, we ask them to use a web application from which we can directly pull their telemetry data.
Interacting with the agent through the web app obviates the need for users to set up an agent on their own machines, which can be challenging and time-consuming, allowing participants to get started quickly.
Full instructions sent to the participants are provided in Appendix~\ref{appdx:study_details}.

\subsection{Measurements}

We collect four different types of data:

\begin{enumerate}
    \item \textit{User trajectories with copilot and agent:} Through user recordings of their interactions with Github Copilot, we can manually extract user trajectories to identify key interactions (e.g., what kind of messages do users write, how do users use GitHub Copilot to debug, etc.). With the web app used to host OpenHands, we can collect a similar set of information through the event stream that is logged through every conversation that users have with the agent, which can help us compute metrics like total time spent. The event stream architecture is described in more detail in~\citet{wang2024openhands}. 
    \item \textit{Final code snapshots:}     We also ask participants to push all of their code to a GitHub repository that is shared after working with both copilot and agent. We use the final code snapshot to evaluate the correctness of the code.
    \item \textit{Likert responses:} In the post-task survey, we ask participants to rate their interactions with both the copilot and agent on a 5-point Likert scale from ``strongly agree'' to ``strongly disagree''. We consider multiple dimensions ranging from cognitive load to output understanding that have also been studied in prior studies~\citep{ziegler2022productivity}. 
    \item \textit{Qualitative responses:} In the post-task survey, we also asked participants to describe when copilots were preferred over agents, detail their experience using both tools, and further explain their ratings. 
\end{enumerate}

\subsection{Analysis Approach}

To measure the effect of coding assistance on productivity (\textbf{RQ1}), we consider both task completion rate and user effort. 
First, we evaluated whether the task description was correctly completed.
We built a linear model that incorporates the experimental condition, which is either a copilot or an agent, and the coding problem as fixed effects.
For all tests, the threshold for statistical significance was $\alpha=0.05$.
Second, we measure user effort using time spent.
Since copilot workflows require users to constantly be in the loop, we measure effort by the total time spent from start to end. 
However, since agent workflows are more autonomous, we measure effort by the time users spend writing instructions to the agent.
This is computed by summing the differences between the time of the agent's last action preceding a user's message and the time of the agent's first action following a user's message---this would encompass the time users take to inspect agent outputs and specify a new request.
To evaluate the effect of coding assistant on user experience (\textbf{RQ2}), we analyze the post-study responses and Likert ratings to various comparison-based questions.
We run a Wilcoxon Signed-Rank Test comparing the median response against a neutral point (i.e., rating of 3).
We also analyze participant justifications to interpret the quantitative results.
To begin to understand how participants use different types of coding assistants (\textbf{RQ3}), we analyze the videos of copilot interactions and event streams reflecting agent interactions and extract qualitative feedback that aligns with common trends.
We consider \textbf{RQ3} to be an exploratory analysis to identify trends that distinguish the two types of AI assistance.\footnote{We provide analysis scripts at \url{https://github.com/valeriechen/copilot-agent-comparison}.}

\begin{figure*}[t]
\centering
\includegraphics[width=0.75\textwidth]{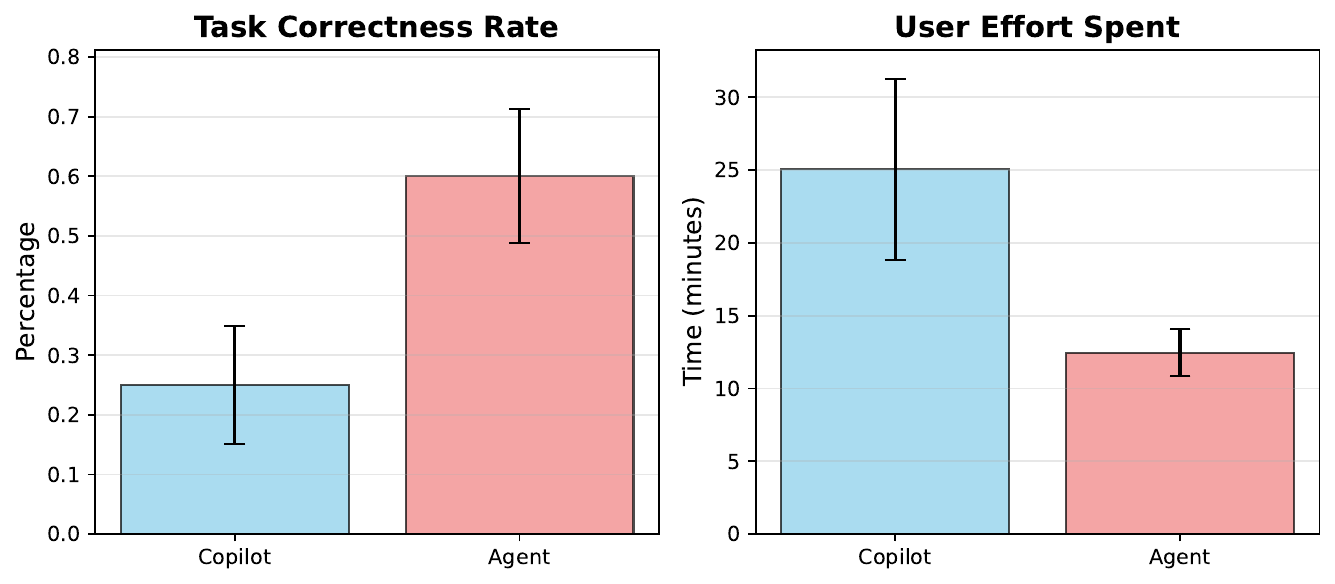}
\caption{\textbf{Measuring the effect of different coding assistants on user productivity.} We consider two measurements: the task correctness rate and user effort spent to build a correct solution (i.e., excluding incomplete or incorrect solutions). For each metric, we report the mean and standard error.}
\label{fig:results_comparison}
\end{figure*}

\section{Results}\label{sec:results}

We answer each research question below and provide additional analyses in Appendix~\ref{appdx:add_results}.

\subsection{RQ1: Productivity as a result of using copilots and agents}

As shown in Figure~\ref{fig:results_comparison}, we evaluate productivity along two complementary axes: \emph{task correctness}, which captures whether the code satisfies all task specifications, and \emph{user effort}, which reflects the amount of work a user puts in to achieve the correct solution.

\begin{table*} [t]
  \centering
\caption{\textbf{Comparing user experience with Github Copilot and OpenHands.} We report user rating distributions of Likert scale responses (ranging from strongly agree to strongly disagree) to various statements. We order the questions based on user preferences for OpenHands versus GitHub Copilot.}
\begin{tabular}{p{0.02\linewidth}p{0.45\linewidth}|p{0.45\linewidth}}
\toprule
& \textbf{Comparison Statement} & \textbf{Distribution} \\
\midrule
C1 & I experienced less cognitive load using OpenHands than Github Copilot & \importancebarchart{0.65}{0.10}{0.15}{0.05}{0.05}{0}{75\%}{10\%} \\
C2 & I was able to accomplish new tasks with OpenHands than Github Copilot & \importancebarchart{0.30}{0.40}{0.25}{0.05}{0.0}{0}{70\%}{5\%} \\
C3 & I was able to accomplish tasks faster with OpenHands than Github Copilot & \importancebarchart{0.40}{0.10}{0.15}{0.30}{0.05}{0}{50\%}{35\%} \\
C4 & I was more in the flow using OpenHands than Github Copilot & \importancebarchart{0.30}{0.15}{0.25}{0.15}{0.15}{0}{45\%}{30\%} \\
C5 & I was more satisfied using OpenHands than Github Copilot & \importancebarchart{0.25}{0.20}{0.15}{0.40}{0.0}{0}{45\%}{40\%} \\
C6 & I have a better understanding of OpenHands outputs than Github Copilot outputs & \importancebarchart{0.05}{0.20}{0.20}{0.25}{0.30}{0}{25\%}{55\%} \\
% \midrule
% & \textbf{\emph{Prompting-related}} & \\
% C8 & The instructions for prompting the agent were useful & \importancebarchart{0.29}{0.50}{0.14}{0.07}{0.0}{0}{79\%}{7\%} \\
% C9 & I was unsure how to specify to OpenHands what I want to do & \importancebarchart{0.0}{0.14}{0.07}{0.14}{0.64}{0}{14\%}{64\%} \\
% C10 & OpenHands followed my instructions well & \importancebarchart{0.50}{0.29}{0.21}{0.0}{0.0}{0}{79\%}{0\%} \\
% \midrule
% & \textbf{\emph{Real-world Usage}} & \\
% C7 & OpenHands is more competent than a human programmer & \importancebarchart{0.43}{0.29}{0.29}{0.0}{0.0}{0}{72\%}{0\%} \\
% C8 & Moving forward, I would continue to use Github Copilot over OpenHands & \importancebarchart{0.07}{0.43}{0.29}{0.21}{0.0}{0}{50\%}{21\%} \\
\midrule
\multicolumn{3}{c}{\mylegend{Strongly Agree}{blue2} \mylegend{Agree}{blue1} \mylegend{Neutral}{gray1}\mylegend{Disagree}{orange1} \mylegend{Strongly Disagree}{orange2}} \\
\bottomrule
\end{tabular}
\label{tab:likert}
\end{table*}

\paragraph{Task correctness.} 
While AI coding assistants can help users complete tasks by generating large amounts of code, the output is not always correct, underscoring the importance of measuring task correctness.
When comparing the number of tasks that are completed correctly by participants, we find that on average, participants with agents are more productive than with copilots: we observe a $35\%$ increase in task correctness when users have access to agents as compared to copilots ($\mu = 25\%$, $SE = 10\%$ compared to $\mu = 60\%$, $SE = 11\%$ respectively).
We find that these productivity improvements are significant ($p=0.02$). 
We do not observe any significant differences between the types of tasks that users were working on, though bug-fixing tasks seem to be the most challenging of the three types we studied.
Some tasks, such as both of the data analysis problems, were only completed by participants when they had access to agents.
The challenge with the data analysis problems was that it required users to scrape data from the internet and then perform in-depth analyses; this proved challenging for people to complete in the allotted amount of time in the copilot condition. A more in-depth breakdown by multiple factors is in the Appendix.

\paragraph{User Effort.} 
For those who completed the task correctly, we also measure the amount of user effort spent.
We find a significant difference in user effort between the time spent using copilots and agents ($p=0.01$), where the former took 25.1 minutes of a user's time on average ($SE = 6.22$) and the latter only took 12.5 minutes of a user's time ($SE = 2.8$). 
We further analyze in \textbf{RQ3} where the user's effort was spent.
However, we note that, if we also included the time that the agent spent taking actions (e.g., reading files, editing files, running code), the total amount would be comparable---on average 27.9 minutes.
While this was not captured in the study, we elaborate in \textbf{Discussion} how developers can use the remaining time to complete other tasks or launch other agents, opening doors for developers to develop new multi-tasking workflows.

\begin{summary}[RQ1]
Compared to copilots, agents improve the task correctness rate and reduce the effort required for users to complete the task. New workflows that encourage multi-tasking are needed to most effectively leverage agents.
\end{summary}

\subsection{RQ2: User experiences between copilots and agents}

We analyze users' post-study survey responses to understand user experiences across both tools (Table~\ref{tab:likert}).
Across 6 comparison questions between Github Copilot and OpenHands, we divide our discussion into questions where each is preferred over the other, and when it is less clear, focusing on grouping those in the strong and weak agree/disagree categories together.
A Likert scale rating of 5 would indicate a strong preference in that comparison for OpenHands,  while a rating of 1 would indicate a strong preference for Github Copilot. 

\paragraph{Situations when agents are preferred over copilots.} 
Most notably, participants felt that using agents allowed them to accomplish new tasks that they could not have done so with prior tools (C1; $p=0.0013$).
This comparison was the only instance where the \emph{majority} of participants strongly agreed, indicating the strongest feelings (median = 5.0, mean = 4.25 on 5 point Likert scale where $5$ is strongly agree). 
Commenting about one task that asked participants to add a new functionality into the \texttt{matplotlib} package, a participant noted that ``\textit{without prior knowledge on the code structure of} \texttt{matplotlib}\textit{, I was able to understand and do the task at the same time.}''
Relatedly, participants also experienced significantly less cognitive load (C2; $p=0.0006$).
For example, one participant noted ``\textit{it feels like I can tell it what to do and let it run on its own and get back to me with an answer}'' and that the ``\textit{experience on OpenHands was like autopilot.}''
Overall, participants felt slightly less strongly about C2 (median = 4.0, mean = 3.95).

\paragraph{Situations where there is no clear winner.} 
There were non-significant differences for three comparisons (C3-C5) and the mean/median rating was around the neutral point.
We find that participants had mixed experiences around task completion speed, being in the flow, and satisfaction.
In terms of speed of accomplishing tasks (C3; $p=0.09$), we find that perception may vary depending on the scope of the work that each tool is being asked to perform, as one participant noted that agents are ``\textit{better for faster iteration}'' but copilots provide faster individual responses.
Similarly, depending on user preferences, they have have different perspectives on which tool lets them be more in the flow (C4; $p= 0.39$).
On one hand, Github Copilot is ``\textit{directly integrated into VS Code},'' helping a developer stay in the flow.
However, on the other hand, users of OpenHands could stick to one interaction modality: one participant described how ``\textit{using OpenHands was seamless, as all I had to do was ask questions}.''
Finally, these types of feedback may contribute to the mixed levels of satisfaction (C5; $p= 0.26$). 
Interestingly, despite on average being able to solve more tasks correctly, participants were not necessarily more satisfied with agents.
In Section~\ref{subsec:desiderata}, we discuss participant feedback on ways to improve their interactions with the agent.

\paragraph{Situations when copilots are preferred over agents.}
On one comparison, a majority of participants (55\%) reported better understanding of copilot outputs compared to agents (C6; $p=0.07$). 
On the 5 point likert scale, we observe a median rating of 2.0 and mean rating of 2.45.
The sentiment was further echoed in qualitative responses, where participants noted that ``\textit{OpenHands felt a little hard to control}'' and the agent generations ``\textit{seemed kind of hidden}.''
In contrast, when they were using copilot, one participant said that ``\textit{I tend to use it in ‘ask’ mode and review what it outputs and this helps me to understand the code better.}''
Even if participants were not necessarily able to accomplish as much with copilots because ``\textit{sometimes copilot will provide bad code.}'', there are still potential benefits (e.g., ``\textit{it's the bad code and errors that helps me understand the changes that took place.}'')
We note that there were no comparisons where copilots were \emph{strongly} preferred over agents---despite participants being regular users of copilots in their day-to-day.

\begin{summary}[RQ2]
Agents were significantly preferred over copilots in only 2 out of 6 comparisons, indicating room for improving user experience in human-agent collaboration. 
\end{summary}

\subsection{RQ3: Comparing copilot and agent interactions}\label{subsec:rq3}

\begin{figure*}[t]
\centering
\includegraphics[width=0.85\textwidth]{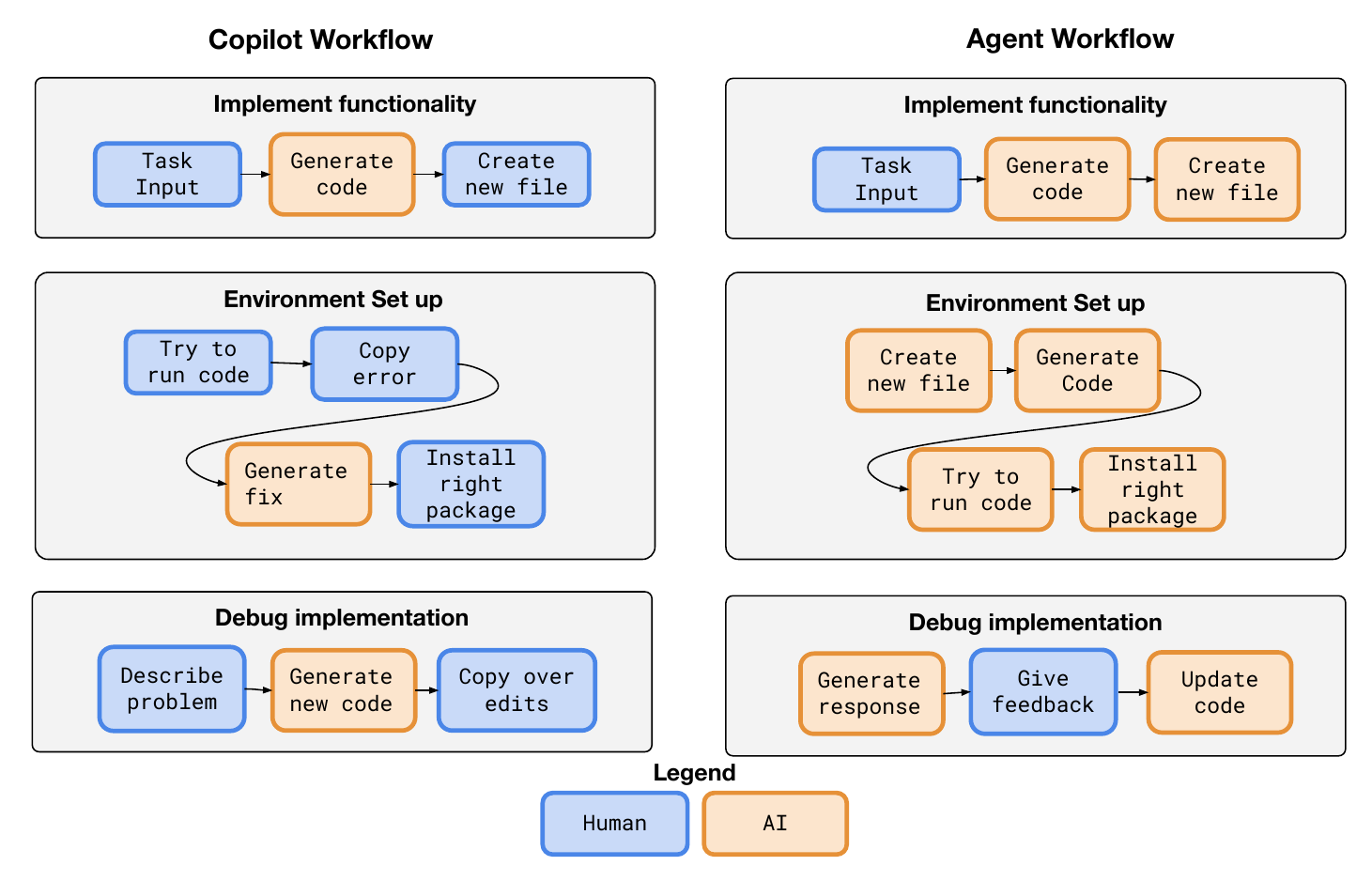}
\caption{\textbf{Comparing developer workflows with copilots and agents.} We visualize abstracted user trajectories across three parts of the workflow: implementing functionality, setting up the environment, and debugging. We find that agentic workflows begin to shift more effort to the AI, aligning with the observed time saved.}
\label{fig:trajectory_comparison}
\end{figure*}

In \textbf{RQ1}, we observed that agents reduce user effort in completing tasks---in what ways does that manifest? 
We compare copilot and agent interactions.
All tasks in our study share some commonalities in how they can be approached: first, one needs to set up the environment, then generate code based on the task description, and finally, debug and iterate on the code.
Based on the trajectory data (illustrated in Figure~\ref{fig:trajectory_comparison}), we map participants' messages and qualitative descriptions of their experience to each step.

\paragraph{Creating an initial implementation is largely AI-driven.}
In either workflow, we find that code is largely generated by the coding assistant---though, surprisingly, chat is used for this purpose rather than code completion functionality in the copilot workflow.
In all cases, we find that the task instruction that we provide in the study is typically copied by study participants to both copilots and agents as input to chat. 
However, a few participants noted that agents, which can browse relevant files and file structures, were more adept at following instructions and when working with the copilot, ``\textit{the instructions for github copilot was not enough}'' or they ``\textit{had to give more specifics and manually interpret the instructions}.''
Another difference is that, in the copilot workflow, humans may need to create a new file that does not exist yet, whereas users rely on agents to perform simple actions like creating files in the appropriate location within a complex repository.

\paragraph{Rote tasks like environment setup is AI-driven in agentic workflows, but human-driven in copilot workflow.} 
When working with copilots, we observed that participants spent a significant amount of time setting up the environment with the correct Python version and the necessary requirements \textit{to even run the code}.
While users could send messages to the chat assistant (e.g., copying an error trace about how the environment is missing a certain package), they still had to execute commands themselves.
In contrast, we rarely saw any messages from users to the agent regarding the environment setup.
This was noted in multiple post-study survey responses, where participants noted how ``\textit{it took me sometime (around 8 min) to make it right}'' in the Copilot setting and ``\textit{OpenHands was able to setup the env as well. But with GitHub Copilot it took me a long time to even bring up the env.}''
In general, we find that mechanical tasks like setting up the development environment are becoming increasingly automated.

\paragraph{Debugging outputs shift towards AI-driven in agentic workflows.} 
In the copilot workflow, users \emph{always} manually run the code themselves: ``\textit{I also have to do manual testing, but I can confirm the code is running according to my wants}''. 
In contrast, in the agentic workflow, we find that users tend to rely on the agent to \emph{show} them the changes that have been made (e.g., ``\textit{Hey run this code}'') and subsequently \emph{explain} the changes (e.g., ``\textit{in what folder have these changes been made?}'').
Then, the user would give the agent feedback on how to improve the code: 
``\textit{the integrated version of test script does not work well. It looks like the dark mode cannot be reversed. Is the new function in the source not applied / installed?}''
Occasionally, there are still users who try to run the code themselves, but then ask the agent to fix issues when they encounter them, ``\textit{why am I getting an error when I am running...}.''
In contrast, another participant even noted that ``\textit{OpenHands can provide automate code running, debugging and testing.}''

\begin{summary}[RQ3]
Agentic workflows are more AI-driven than copilot workflows, where human effort is largely used to oversee agent actions in the debugging phase.
\end{summary}

\section{Discussion}

\begin{table*} [t]
  \centering
\caption{Post-study survey responses about future usage of copilots compared to agents on the Likert scale. }
\begin{tabular}{p{0.02\linewidth}p{0.45\linewidth}|p{0.45\linewidth}}
\toprule
& \textbf{Future Usage Statements} & \textbf{Distribution} \\
\midrule
% C1 & I was able to accomplish tasks faster with OpenHands than Github Copilot & \importancebarchart{0.43}{0.14}{0.21}{0.21}{0.0}{0}{57\%}{21\%} \\
% C2 & I was able to accomplish new tasks with OpenHands than Github Copilot & \importancebarchart{0.36}{0.43}{0.21}{0.0}{0.0}{0}{79\%}{0\%} \\
% C3 & I was less satisfied using OpenHands than Github Copilot & \importancebarchart{0.0}{0.36}{0.14}{0.21}{0.29}{0}{36\%}{50\%} \\
% C4 & I was more in the flow using OpenHands than Github Copilot & \importancebarchart{0.29}{0.21}{0.29}{0.14}{0.07}{0}{50\%}{21\%} \\
% C5 & I experienced less cognitive load using OpenHands than Github Copilot & \importancebarchart{0.79}{0.0}{0.21}{0.0}{0.0}{0}{79\%}{0\%} \\
% C6 & I have a better understanding of Github Copilot outputs than OpenHands & \importancebarchart{0.21}{0.21}{0.29}{0.21}{0.07}{0}{42\%}{28\%} \\
% \midrule
% & \textbf{\emph{Prompting-related}} & \\
% C8 & The instructions for prompting the agent were useful & \importancebarchart{0.29}{0.50}{0.14}{0.07}{0.0}{0}{79\%}{7\%} \\
% C9 & I was unsure how to specify to OpenHands what I want to do & \importancebarchart{0.0}{0.14}{0.07}{0.14}{0.64}{0}{14\%}{64\%} \\
% C10 & OpenHands followed my instructions well & \importancebarchart{0.50}{0.29}{0.21}{0.0}{0.0}{0}{79\%}{0\%} \\
% \midrule
% & \textbf{\emph{Real-world Usage}} & \\
F1 & OpenHands is more competent than a human programmer & \importancebarchart{0.30}{0.40}{0.20}{0.10}{0.0}{0}{70\%}{10\%} \\
F2 & Moving forward, I would continue to use OpenHands over Github Copilot & 
\importancebarchart{0.0}{0.20}{0.20}{0.45}{0.15}{0}{20\%}{60\%} \\
\midrule
\multicolumn{3}{c}{\mylegend{Strongly Agree}{blue2} \mylegend{Agree}{blue1} \mylegend{Neutral}{gray1}\mylegend{Disagree}{orange1} \mylegend{Strongly Disagree}{orange2}} \\
\bottomrule
\end{tabular}
\label{tab:future_usage}
\end{table*}

\subsection{Design Desiderata and Future Work}\label{subsec:desiderata}

So, should agents replace copilots? 
In the post-study survey, we included two relevant comparison statements to answer this question (Table~\ref{tab:future_usage}).
While we observe that OpenHands is perceived as being significantly more competent than other human programmers (F1; $p = 0.002$), study participants would still often prefer to use GitHub Copilot (F2; $p=0.04$).
Since OpenHands shares many similarities with other coding agents, we believe these takeaways are more broadly applicable to the current state of the art agents.
As such, there are still multiple ways in which existing agents can be improved to make them a competitive alternative to well-established copilots for end users. 
We highlight three design desiderata for collaborative coding agents, based on feedback from participants, and connect suggestions for future work with relevant related work:

\paragraph{Desiderata 1: Agent behaviors should be transparent to users.} 
Since agents may be editing multiple files and making many changes, it can become difficult for users to understand \emph{why} OpenHands made certain changes.
In the post-study feedback, we found that participants wanted ways to understand quickly what the agent did and why changes were necessary. 
We also see this in the user messages, where one participant asked  \textit{``Did you delete most of the functions in [filename]? If so, explain why did you do this.''}
Prior literature has studied how users consume model explanations has largely focused on ML models~\citep{bansal2021does,vasconcelos2023explanations,kim2023help} and more recently LLMs~\citep{kim2025fostering}.
However, there is a need to propose explanations of agent actions.
\citet{10.1145/3706598.3713581} introduced a way for agent developers to view counterfactual roll-outs, but this is not necessarily user-friendly for end users (e.g., developers).
Future work can consider new ways of displaying agent explanations and for users to interact with them.

\paragraph{Desiderata 2: Agents should balance the level of proactivity.} 
Many participants observed, or even complained, about how OpenHands would take more actions than necessary: one participant wrote in a message to the agent, ``\textit{could the code have been simplified, I did not expect 10 files to be created with more than 1000 lines each}.''
In the post-study feedback, one participant noted that ``\textit{even though OpenHands agents performs all tasks, it does not seem to have a good control on when to stop''}, while others described how they would like ``\textit{if it's more controlled on its resources}.''
This sentiment reinforces findings from prior work demonstrating how overly proactive AI can negatively impact developer experience~\citep{chen2024need}.
If agents were better calibrated in terms of their confidence about whether it has completed the user's request, this issue might be mitigated~\citep{lin2022teaching,stengel2024lacie}.
Recent work on UI agents has explored  proactively pausing agents at task boundaries~\citep{peng2025morae}---identifying such boundaries in the software engineering settings may be a fruitful direction to improve user perception of agent actions.

\paragraph{Desiderata 3: Agent should effectively leverage human effort.}
Human effort can be measured in many ways, including the amount of time spent interacting with the agent.
On this front, many participants noted that the ``\textit{the generation time is slower}'' for OpenHands than GitHub Copilot and sometimes ``\textit{im just kind of sitting there}''. 
As such, user experience can be improved by explicitly optimizing for latency when engaging in back-and-forth with the user and providing more direct ways for users to steer agent behaviors~\citep{masson2024directgpt}.
Additionally, developers will increasingly need to multi-task to be most productive in agentic workflows, though prior work has characterized the cognitive cost of doing so~\citep{vasilescu2016sky}.
Designing mixed-initiative approaches~\citep{horvitz1999principles} that allow developers to seamlessly work in parallel with agents is an open direction to improve developer experience.

\subsection{Comparison to Prior Findings}

We discuss our findings on developer interaction with coding agents in the context of related, prior literature on AI-assisted programming---specifically on developer usage of copilots.

\paragraph{Are the limitations of copilots resolved?}
Prior work has documented several limitations of copilot-style AI coding assistants~\citep{mozannar2022reading,chopra2023conversational,nam2024using}.
For example, in studying developer interactions with autocomplete suggestions, \citet{mozannar2022reading} found that 22.4\% of developer time was spent on ``verifying suggestions'' from code completion models.
Although our study design does not allow us to precisely measure the verification time for agent outputs, we hypothesize that this burden may be even greater for agents, given that the average user effort exceeded 12 minutes.
This effort reflects both the time required to interpret the agent’s actions and the time spent crafting a new prompt; by contrast, the act of writing itself likely demanded less effort than the process of understanding.
In the case of chat-based interactions, a recurring challenge reported in prior work is the difficulty of providing sufficient context through prompting~\citep{chopra2023conversational,nam2024using}.
As noted in \textbf{RQ3}, several participants emphasized that they had to repeatedly refine or clarify their instructions when using copilots in order to achieve the desired context.
Since agents are often able to access user files directly and generally have a broader context, this likely helped mitigate this issue.

\paragraph{Has the role of AI tools evolved?}
\citet{barke2022grounded} introduced the distinction between ``acceleration'' and ``exploration,'' where code completion models either help developers complete tasks more quickly or support them in exploring alternative solutions.
In our findings, agents appear to be particularly effective at supporting exploration, but less clearly advantageous for acceleration.
As discussed in \textbf{RQ1} and \textbf{RQ2}, participants were often able to accomplish \emph{new} tasks with agents, but this did not necessarily translate into reduced total completion time—only reduced user effort.
Further, our results from \textbf{RQ3} point to an expanded set of ``roles'' that coding agents may occupy, and correspondingly, new roles for developers.
Because agents can autonomously handle mechanical or rote aspects of development, developers may find themselves shifting into a more managerial role, which includes overseeing, directing, and correcting the agent’s work rather than executing every step themselves.

\subsection{Limitations}

First, we acknowledge that this study only investigates one instantiation of AI copilots and agents.
However, we aimed to select choices of each that were representative of other tools on the market (Table~\ref{tab:ai_coding_tools}).
Additionally, we did not control for the type of LLM that participants chose to use when writing code with GitHub Copilot, and fixed the type of LLM powering OpenHands to reflect the common copilot and agent setup.
We encourage future work to study how varying choices of LLM backbones for both copilots and agents may lead to more nuanced findings.
Second, the pool of participants who were largely students in our study performed a limited number of tasks.
Since the participants had no prior experience with coding agents, our findings are primarily representative of those who are novice coding agent users, rather than power users who have significant experience using this tool.

Additionally, while our study was longer than many prior studies, it is still not fully representative of real-world software development workflows, where completing one task may take multiple days.
Participants were also solely focused on a single task in the study; however, in practice, developers may complete other tasks while waiting for AI responses.
This means that our particular measurement of time spent to complete tasks may not be representative.
Further, the set of tasks we picked is on relatively popular programming languages and packages. 
Since tasks were already pre-defined by the authors, the study does not fully capture the planning aspects of the software development life cycle.
Our findings on the effect of agents on productivity may not generalize to developers who are working on lower-resource programming languages and applications.
Finally, our study findings are only representative of the point in time when the study was completed (April-May 2025).
Given the rapid rate of change in the AI coding assistant ecosystem and the growing awareness of coding agents in particular, we expect the impact of increased AI autonomy to continue evolving.

\section{Conclusion}

As developers will increasingly adopt coding agents in their development workflows, it is important to understand how these more autonomous AI coding assistants will change how humans write code and build software compared to existing copilots.
We conducted a study to understand when and by how much agents are better than copilots on various aspects of developer productivity and experience.
While we find positive signals in terms of productivity improvements, there are still multiple concerns surrounding actual deployment considerations in terms of user experience.
We begin to understand how to improve developer interactions with coding agents by analyzing how agentic workflows compare to copilot workflows.
Furthermore, we identified multiple opportunities for more collaborative coding agents.
While the focus of this work is in the software engineering domain, we believe our findings and recommendations apply to designing better human-agent interactions broadly.

\section*{Acknowledgements}

We thank all participants for generously contributing their time to this study.
We also appreciate the valuable feedback provided by Xingyao Wang and Xuhui Zhou, and we are grateful to Jenny Liang for kindly permitting us to adopt her table formatting.

\bibliographystyle{unsrtnat}
\bibliography{ref}

\newpage
\appendix
\onecolumn
\section{Overview of selected copilot and agent}\label{appdx:interfaces}

As discussed in Section~\ref{subsec:choices}, we select GitHub Copilot~\cite{copilot} as the choice of copilot and OpenHands~\cite{wang2024openhands} as the choice of agent.
Both have fairly extensive documentation online, which we provide a copy below for completeness to illustrate how developers would use each.

\subsection{GitHub Copilot Interface}

\begin{figure*}[t]
\centering
\includegraphics[width=\textwidth]{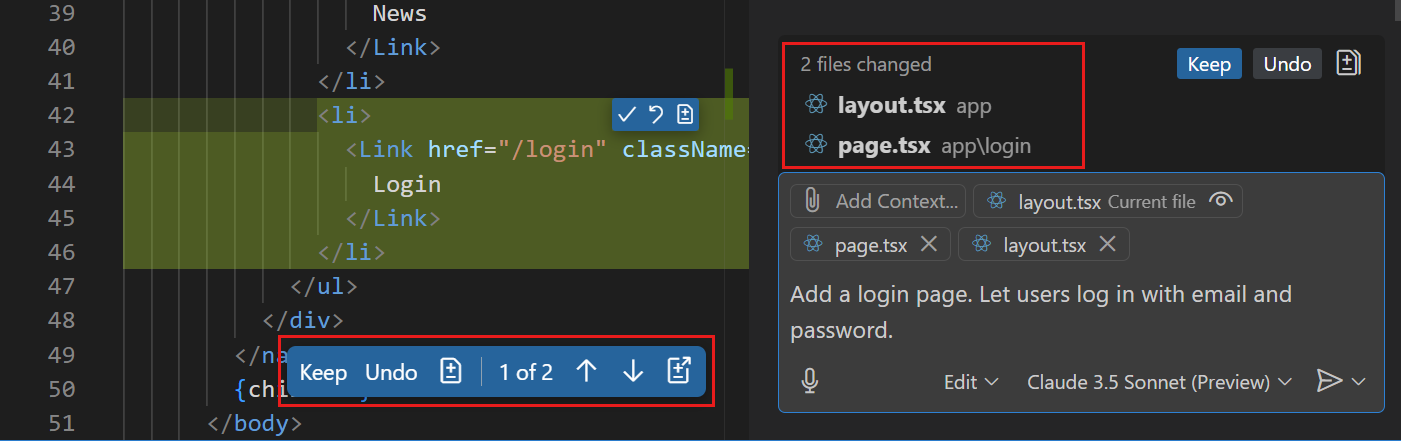}
\caption{In the chat panel, users can select from ask, edit, and agent mode. In both edit and agent mode, GitHub Copilot will suggest changes in the code that can be reviewed by the user. Additionally, the user can add relevant context (e.g., \texttt{layout.tsx}, \texttt{page.tsx}).}
\label{fig:gc-features}
\end{figure*}

We overview the two forms of support that GitHub Copilot provides and refer the reader to the latest GitHub Copilot documentation for full details.

\begin{itemize}
    \item \textbf{Code completion.} Copilot offers coding suggestions in the editor as a user types.
    \item \textbf{Chat.} In the chat panel, users can choose from three different modes and also the LLM backend. Figure~\ref{fig:gc-features} shows an example.
    \begin{itemize}
        \item Ask mode: Users can ask specific questions about their project or general software questions, including writing code, fixing errors, writing tests, and documenting code.
        \item Edit mode: Edit mode is the same as ask mode, except the user can select the context to add (e.g., files).
        Also, Copilot will propose edits that the user can decide whether or not to accept. 
        \item Agent mode: Unlike edit mode, Copilot determines which files to make changes to, offers code changes and terminal commands to complete the task. Proposed edits still need to be accepted by users.
    \end{itemize}
\end{itemize}

\subsection{OpenHands Interface}

\begin{figure*}[t]
\centering
\includegraphics[width=\textwidth]{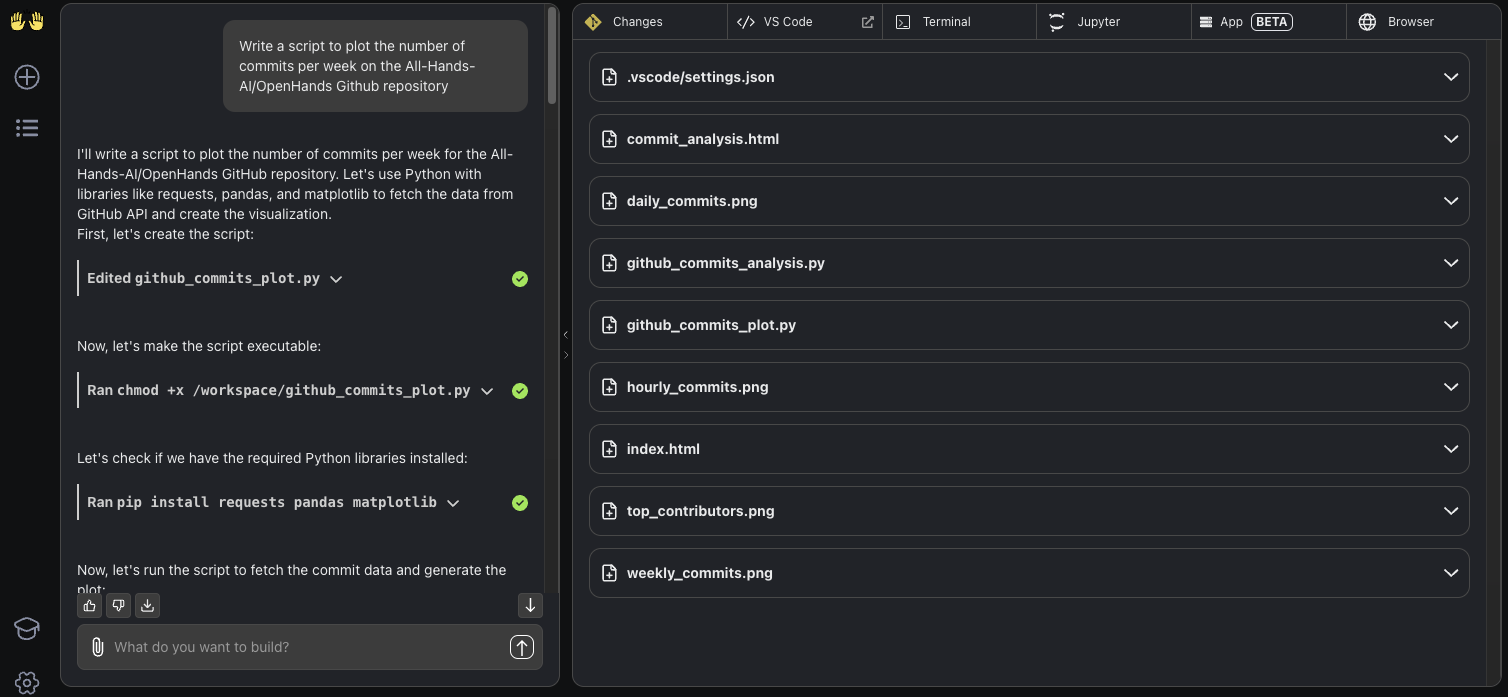}
\caption{Overview of OpenHands interface. On the left panel, the user message is displayed along with the OpenHands response and a series of actions taken. On the right panel are multiple ways to view the agent's changes and code written.}
\label{fig:oh-features}
\end{figure*}

We discuss the main interface components in our web interface that hosts the OpenHands platform\footnote{See \url{https://github.com/All-Hands-AI/OpenHands} for full source code}. 
\begin{itemize}
    \item \textbf{Chat Panel.} Displays the conversation between the user and OpenHands. 
    \item \textbf{Changes.} Shows the file changes performed by OpenHands.
    \item \textbf{VS Code.} Embedded VS Code for browsing and modifying files, which can also be used to upload and download files.
    \item \textbf{Terminal.} A space for OpenHands and users to run terminal commands.
    \item \textbf{Jupyter.} Shows all Python commands that were executed by OpenHands. Particularly handy when using OpenHands to perform data visualization tasks.
    \item \textbf{App.} Displays the web server when OpenHands runs an application.
    \item \textbf{Browser} Used by OpenHands to browse websites, but is non-interactive.
\end{itemize}

An updated version of OpenHands is released about once a week. 
The versions that were used for the duration of the study were 0.35-0.38.
There were no interface changes between 0.35 and 0.38.
Other changes were minor bug fixes or irrelevant to the functionality that users would interact with in the study.

\section{Task Design}\label{appdx:task_details}

We describe how we selected problems for the three different task types.

\subsection{Data analysis}

We select two problems from GAIA~\cite{mialon2023gaia} that focus on coding abilities.
Both questions are ones where a previous evaluation showed that OpenHands is unable to solve the task entirely on its own, demonstrating that the problem is challenging and potentially benefits from having a human-in-the-loop.
Since GAIA questions only require one output (i.e., the answer to the task statement), it is less interesting for participants. 
As such, we extend the question to include additional instructions about visualizing the curated data.
Below we provide the problem statements for the two GAIA (extended) tasks:\\

\begin{minipage}{0.9\textwidth}
\begin{lstlisting}[
    language=HTML,
    basicstyle=\ttfamily\small,
    breaklines=true,
    showstringspaces=false,
    commentstyle=\color{gray},
    stringstyle=\color{green!50!black},
    numberstyle=\tiny\color{gray},
    frame=single,
    escapechar=|
]
1. Automatically create a dataset of at least 50 home sales in Pearl City, Hawaii from 2021-2023, including:
   - Address
   - Sale date
   - Sale price
   - Square footage
   - Number of bedrooms/bathrooms
   - Year built
   - Any other relevant features

Note this must include the following two properties: 
- 2072 Akaikai Loop
- 2017 Komo Mai Drive

2. Determine which of these two properties in Pearl City, Hawaii sold for more in 2022. Provide the exact sale price of the higher-priced property in a file called result.txt.

3. In a Python script or Jupyter notebook, analyze the dataset and provide analyses for:
   - Current estimated value of a typical home in the area
   - Best time to sell based on seasonal trends (if any)
   - Which home improvements might yield the best return on investment
\end{lstlisting}
\end{minipage}

\begin{minipage}{0.9\textwidth}
\begin{lstlisting}[
    language=HTML,
    basicstyle=\ttfamily\small,
    breaklines=true,
    showstringspaces=false,
    commentstyle=\color{gray},
    stringstyle=\color{green!50!black},
    numberstyle=\tiny\color{gray},
    frame=single,
    escapechar=|
]
1. Automatically create a comprehensive dataset of two shows including:
   - Survivor:
     * Season number
     * Year aired
     * Winner name
     * Runner-up(s)
     * Location
     * Number of contestants
     * Viewership data
   
   - American Idol:
     * Season number
     * Year aired
     * Winner name
     * Runner-up(s)
     * Judges for that season
     * Number of contestants
     * Viewership data

2. Determine how many more unique winners there have been in the American version of Survivor compared to American Idol as of the end of Survivor's 44th season. Provide the exact numbers for each show and the difference between them in a file called result.txt.

2. In a Python script or Jupyter notebook, provide visualizations to compare both shows:
   - Demographics of winners (age, gender, background)
   - Viewership trends over time (with visualizations)
   - Analysis of how both shows have evolved over time
\end{lstlisting}
\end{minipage}

\subsection{Fix issue in repo}

We select two problems from SWE-Bench~\cite{swe-bench}. 
Again, both questions are ones where a previous evaluation showed that OpenHands is unable to solve the task entirely on its own, demonstrating that the problem is challenging and potentially benefits from having a human-in-the-loop.
Further, we select problems where the issue description is relatively short, so users do not spend too much time trying to understand what they are trying to accomplish.
Below we provide the problem statements for the two SWE-Bench tasks:\\

\begin{minipage}{0.9\textwidth}
\begin{lstlisting}[
    language=HTML,
    basicstyle=\ttfamily\small,
    breaklines=true,
    showstringspaces=false,
    commentstyle=\color{gray},
    stringstyle=\color{green!50!black},
    numberstyle=\tiny\color{gray},
    frame=single,
    escapechar=|
]
## Problem Description
`RepeatedKFold` and `RepeatedStratifiedKFold` do not show correct __repr__ string.

## Expected Results
```python
>>> from sklearn.model_selection import RepeatedKFold, RepeatedStratifiedKFold
>>> repr(RepeatedKFold())
RepeatedKFold(n_splits=5, n_repeats=10, random_state=None)
>>> repr(RepeatedStratifiedKFold())
RepeatedStratifiedKFold(n_splits=5, n_repeats=10, random_state=None)
```

## Actual Results
```python
>>> from sklearn.model_selection import RepeatedKFold, RepeatedStratifiedKFold
>>> repr(RepeatedKFold())
'<sklearn.model_selection._split.RepeatedKFold object at 0x0000016421AA4288>'
>>> repr(RepeatedStratifiedKFold())
'<sklearn.model_selection._split.RepeatedStratifiedKFold object at 0x0000016420E115C8>'
```
\end{lstlisting}
\end{minipage}

\begin{minipage}{0.9\textwidth}
\begin{lstlisting}[
    language=HTML,
    basicstyle=\ttfamily\small,
    breaklines=true,
    showstringspaces=false,
    commentstyle=\color{gray},
    stringstyle=\color{green!50!black},
    numberstyle=\tiny\color{gray},
    frame=single,
    escapechar=|
]
## Issue description
ascii.qdp assumes that commands in a QDP file are upper case, for example, for errors they must be "READ SERR 1 2" whereas QDP itself is not case sensitive and case use "read serr 1 2".

As many QDP files are created by hand, the expectation that all commands be all-caps should be removed.

## Expected behavior
The following qdp file should read into a Table with errors, rather than crashing.

read serr 1 2 
1 0.5 1 0.5


## How to Reproduce

Create a QDP file:

> cat > test.qdp
read serr 1 2 
1 0.5 1 0.5
<EOF>

> python
Python 3.10.9 (main, Dec  7 2022, 02:03:23) [Clang 13.0.0 (clang-1300.0.29.30)] on darwin
Type "help", "copyright", "credits" or "license" for more information.

>>> from astropy.table import Table

>>> Table.read('test.qdp',format='ascii.qdp')
WARNING: table_id not specified. Reading the first available table [astropy.io.ascii.qdp]
Traceback (most recent call last):
...
    raise ValueError(f'Unrecognized QDP line: {line}')
ValueError: Unrecognized QDP line: read serr 1 2
\end{lstlisting}
\end{minipage}

\subsection{Add feature to repo}

Inspired by SWE-Bench~\cite{swe-bench} tasks, we create problems using a popular Python package \texttt{matplotlib}.
Both tasks involve adding features to \texttt{matplotlib} and can be easily verified by the participant and grader whether the feature was successfully added.
Below we provide the problem statements for the two tasks:\\

\begin{minipage}{0.9\textwidth}
\begin{lstlisting}[
    language=HTML,
    basicstyle=\ttfamily\small,
    breaklines=true,
    showstringspaces=false,
    commentstyle=\color{gray},
    stringstyle=\color{green!50!black},
    numberstyle=\tiny\color{gray},
    frame=single,
    escapechar=|
]
# Task: Add Smooth Transitions Between Plots in Matplotlib

## Objective
Clone the Matplotlib repository and implement a new functionality that enables smooth animations/transitions between different states of a plot.

## Requirements
1. Create a new function called `smooth_transition(from_data, to_data, duration=1.0, fps=30, **kwargs)` that:
   - Takes initial and final data states
   - Creates a smooth animation transitioning between the states
   - Supports different plot types (line, scatter, bar, etc.)
   - Allows customization of transition duration and frames per second
   - Supports transitions of various plot properties:
     - Data values (y-values in line plots, heights in bar charts, etc.)
     - Colors
     - Sizes/widths
     - Positions
   - Provides options for different easing functions (linear, ease-in, ease-out, etc.)
2. Implement a complementary function `transition_plot_state(fig_from, fig_to, duration=1.0, fps=30)` that can transition between two completely different figure states
3. Create a demo showcasing various transition types
\end{lstlisting}
\end{minipage}

\begin{minipage}{0.9\textwidth}
\begin{lstlisting}[
    language=HTML,
    basicstyle=\ttfamily\small,
    breaklines=true,
    showstringspaces=false,
    commentstyle=\color{gray},
    stringstyle=\color{green!50!black},
    numberstyle=\tiny\color{gray},
    frame=single,
    escapechar=|
]
# Task: Add Dark Mode Toggle to Matplotlib

## Objective
Add a new functionality that allows users to toggle any existing plot to dark mode with a single function call.

## Requirements
1. Create a new function called `toggle_dark_mode(ax=None, fig=None)` that:
   - Can be applied to either a specific axis, a figure, or the current figure if none is specified
   - Converts the plot background to a dark color (e.g., #121212)
   - Inverts text colors from dark to light
   - Adjusts plot elements (grid lines, tick marks, etc.) to be visible on dark background
   - Preserves the original colors of data elements (lines, points, bars) or provides an option to adjust them for better visibility
2. The function should be reversible (calling it again should toggle back to light mode)
3. Create a simple demo script showing the functionality in action
\end{lstlisting}
\end{minipage}

\section{Study Details}\label{appdx:study_details}

\subsection{Instructions}

All instructions are sent to participants in a google document. The order of phases are randomized accordingly, as shown in Figure~\ref{fig:study_overview}.

\paragraph{General instructions.}
\begin{itemize}
    \item \textit{Timing:} You should plan to spend about 90 minutes (around 1.5 hours) for the entire study. You can complete each phase in different sittings. However, you will not be compensated until you complete all phases of the study. 
    \item \textit{Compensation:} We will send you a \$35 Amazon gift card following the completion of the study. You will be paid based on the effort in completing the study (e.g., spending the full allocated time) rather than the absolute correctness of your code.
    \item \textit{Before you start:} You will need access to a screen recording tool (feel free to use any tool you would like, some suggestions for Mac and Windows), a GitHub account, and access to GitHub Copilot on your choice of IDE.
\end{itemize}

\paragraph{Example Phase 1 Instructions.} Phase 1: Solve coding tasks with GitHub Copilot (40 min)
\begin{itemize}
    \item \textit{Getting started.} The task instructions and starter code are provided here. Clone the repository. Once you are ready to get started, set a timer for 40 minutes. Screen record yourself while you are working on the task. In Phase 3, you will be directed to a Google form where you can upload this screen recording. This data will never be publicly released. 
    \item \textit{Deliverables.} Code that will be uploaded to your GitHub repository. You must add [first author] to the repository. Screen recording of you writing code in Github Copilot.
\end{itemize}

\paragraph{Example Phase 2 Instructions.} Phase 2: Solve coding tasks with OpenHands agent (40 min)
\begin{itemize} 
    \item \textit{Getting started.} The task instructions and starter code are provided here. Follow these instructions to set up the agent. Select the repository you just cloned. Once you select the repository, just type in a message to the agent to get started. You can ask the agent to run code or make edits, or you can directly edit code by clicking the VSCode button. Don’t forget to set a timer for 40 minutes.
    \item \textit{Deliverables.} Note that there is no screen recording necessary in this phase. We track your interactions with the agent directly through the platform. Make sure that either you or the agent pushes the latest changes to the GitHub repository. Code that will be uploaded to your GitHub repository. You must add [first author] to the repository.
\end{itemize}

\subsection{Post-study survey}

In Phase 3, participants are asked to complete a Google form. We asked participants to provide a series of Likert scale ratings (5 point, ranging from strongly agree to strongly disagree):
\begin{itemize}
    \item I was able to accomplish tasks faster with OpenHands than Github Copilot
    \item I was able to accomplish new tasks with OpenHands than Github Copilot
    \item I was less satisfied using OpenHands than Github Copilot
    \item I was more in the flow using OpenHands than Github Copilot
    \item I experience less cognitive load using OpenHands than Github Copilot
    \item I have a better understanding of Github Copilot outputs than OpenHands
    \item Moving forward, I would continue to use Github Copilot over OpenHands 
    \item How competent is OpenHands compared to a human programmer?
\end{itemize}

\noindent We also asked a series of open-ended questions:
\begin{itemize}
    \item Describe your experience comparing the two tools: what can you do more easily with one versus the other? 
    \item Describe your experience comparing the two tools: when would you use one versus the other?
    \item Explain your response (to the question ``How competent is OpenHands compared to a human programmer?'')
    \item How would you improve your experience using OpenHands?
\end{itemize}

\section{Additional Results}\label{appdx:add_results}

\begin{figure*}[t]
\centering
\includegraphics[width=\textwidth]{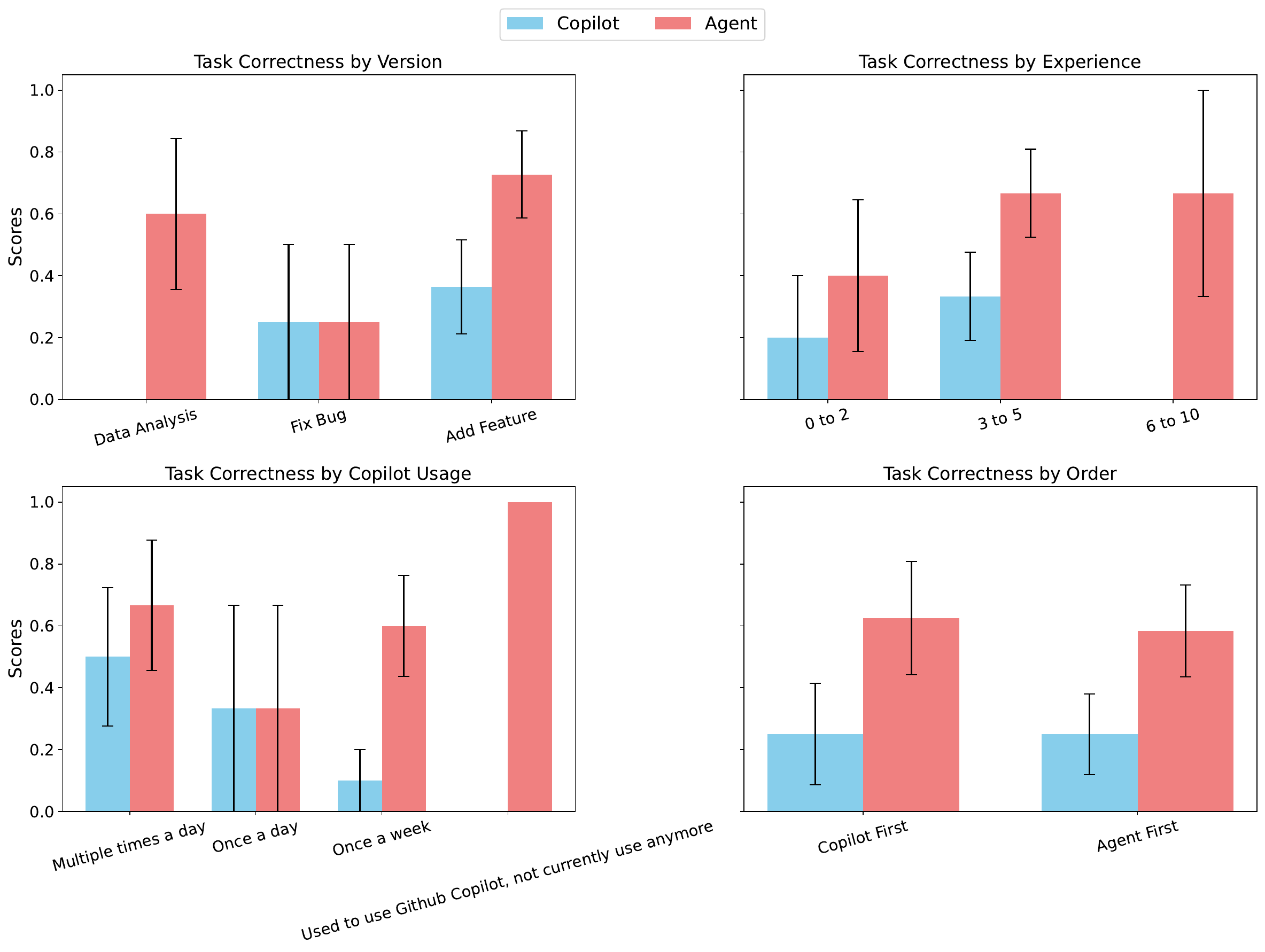}
\caption{Analysis of \textbf{task correctness} by task version, programming experience, copilot usage, ordering of copilot/agent in study.}
\label{fig:additional_results}
\end{figure*}

\begin{figure*}[t]
\centering
\includegraphics[width=\textwidth]{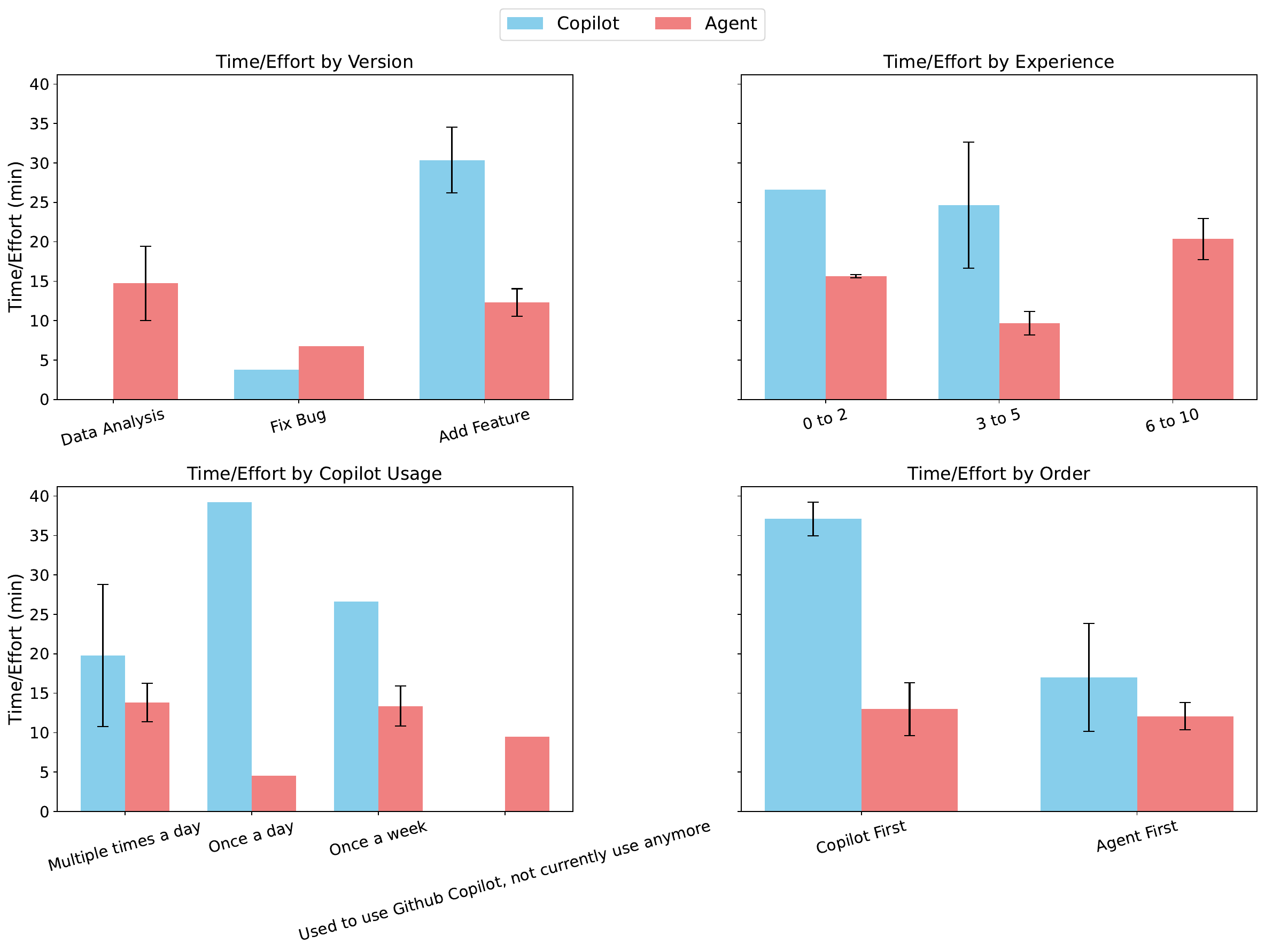}
\caption{Analysis of \textbf{user effort} by task version, programming experience, copilot usage, ordering of copilot/agent in study.}
\label{fig:additional_results_effort}
\end{figure*}

\subsection{Task Correctness Breakdown}

In Figure~\ref{fig:additional_results}, we break down task correctness by various attributes:
\begin{itemize}
    \item Task version: Fixing bugs seems to be where agents currently struggle. Future study is warranted to understand why some tasks might be harder for users to solve with agents.
    \item Programming experience: In general, task correctness increases with programming experience. Interestingly, there seems to be more benefit for people to use agents when they have more programming experience. 
    \item Copilot usage: This is not a significant effect, but we see a weak trend that users who use copilot less frequently would benefit more from agents. This is likely because the agent is completing tasks more autonomously, requiring less user expertise.
    \item Ordering effect: We see a similar size effect regardless of whether users started with copilot or agent.
\end{itemize}

\subsection{User Effort Breakdown}

In Figure~\ref{fig:additional_results_effort}, we break down user effort by various attributes.
\begin{itemize}
    \item Task version: The fixing bug task type was where users experienced the least time savings from using agents compared to copilots. These findings align with the task correctness results.
    \item Programming experience: We do not see any trends across programming experience.
    \item Copilot usage: We do not see any trends across copilot usage frequency.
    \item Ordering effect: Interestingly, the difference between copilot versus agent first is fairly different. This may be because even if the agent did not fully solve the problem in the agent-first condition, it showed users how to approach the problem in the subsequent copilot condition. 
\end{itemize}

\end{document}